# Energy-tunable entangled photon sources on a III-V/Silicon chip

*Yan Chen,[1] Jiaxiang Zhang,[1] Michael Zopf,[1] Kyubong Jung,[1] Yang Zhang,[1] Fei Ding,[1]\* and Oliver G. Schmidt[1,2]*

[1]Institute for Integrative Nanosciences, IFW Dresden, Helmholtzstraße 20, 01069 Dresden, Germany

[2]Material Systems for Nanoelectronics, Chemnitz University of Technology, Reichenhainerstrasse 70, 09107 Chemnitz, Germany

**Corresponding author**:

Fei Ding

Institute for Integrative Nanosciences, IFW Dresden,

Helmholtzstraße 20, 01069 Dresden, Germany

Tel: +49 351 4659 752

Email: f.ding@ifw-dresden.de

**Many of the envisioned quantum photonic technologies, *e.g.* a quantum repeater, rely on an energy- (wavelength-) tunable source of polarization entangled photon pairs.[1, 2, 3] The energy tunability is a fundamental requirement to perform two-photon-interference between different sources and to swap the entanglement.[2] Parametric-down-conversion[4] and four-wave-mixing[5] sources of entangled photons have shown energy tunability,[6] however the probabilistic nature of the sources limits their applications in complex quantum protocols. Here we report a silicon-based hybrid photonic chip where energy-tunable polarization entangled photons are generated by deterministic and scalable III-V quantum light sources. This device is based on a micro-electromechanical system (MEMS) incorporating InAs/GaAs quantum dots (QDs) on a PMNPT-on-silicon substrate. The entangled photon emissions from single QDs can be tuned by more than 3000 times of the radiative linewidth without spoiling the entanglement. With a footprint of several hundred microns, our design facilitates the miniaturization and scalable integration of indistinguishable entangled photon sources on silicon. When interfaced with silicon-based quantum photonic circuits, this device will offer a vast range of exciting possibilities.[7]**

A topical challenge in experimental quantum photonics is the generation and manipulation of polarization entangled photon pairs.[8, 9] Spontaneous parametric-down-conversion (SPDC) and four-wave-mixing (FWM) have served as the main workhorses for these purposes in the past decade, and the implementation of a fully integrated quantum optoelectronic device is within reach by marrying these sources with chip-scale silicon photonics.[7, 10, 11, 12] However the generated photons are characterized by Poissonian statistics, *i.e.* one usually does not know when an entangled photon pair is emitted. This fundamentally limits their applications in complex quantum protocols, *e.g.* an event-ready test of Bell's inequality and high efficiency entanglement purifications, where deterministic operations are much favored.[8]

The intrinsic limitations of SPDC and FWM processes call for next generation entangled photon sources. III-V semiconductor quantum dots, often referred to as artificial atoms, are among the leading candidates for deterministic quantum light sources. As proposed by Benson *et al.* single QDs can generate polarization entangled photon pairs via its biexciton (XX) cascade decay through the intermediate exciton states X, see Fig. 1.[13] In real III-V QDs the anisotropy in strain, composition and shape reduces the QD symmetry to $C_{2v}$ or the even lower $C_1$, leading to the appearance of an energetic splitting between the two bright X states, the so-called fine structure splitting (FSS).[14] High fidelity to the entangled state $|\Psi^+\rangle = 1/\sqrt{2}\,(|H_{XX}H_X\rangle + |V_{XX}V_X\rangle)$, with H and V denoting the horizontal and vertical polarizations, can be observed only with a vanishing FSS (typically, smaller than the radiative linewidth of ~ 1µeV). The probability of finding such QDs in an as-grown sample is $< 10^{-2}$. After extensive efforts by many groups, the elimination of FSS can be achieved by applying rapid thermal annealing,[15] optical Stark effect,[16] magnetic field,[17, 18] electric field,[19, 20] and more recently, anisotropic strain fields[21, 22] to the QDs. In the past years we have witnessed considerable progress in this field, and entangled photon emissions can be triggered optically[17, 23] from single QDs with high brightness (up to 0.12 pair per excitation pulse)[24] and high indistinguishability (0.86 ± 0.03 for the XX photons).[25] III-V QDs also possess an important advantage of being compatible with mature semiconductor technology, and electrically triggered entangled photon sources have been successfully demonstrated.[22, 26]

Armed with these powerful techniques, III-V QDs have the potential to fulfill the "wish-list" of a perfect entangled photon source.[27] Among the next goals are the miniaturization and scaling up of the technology. However, serious problems exist. *First*, the FSS of each QD can only be eliminated under particular tuning parameters, and any attempt to manipulate the emission energy increases the FSS and spoil the entanglement. This fact undoubtedly restricts the entangled photon emissions at single/arbitrary energies. *Second*, as being investigated with SPDC and FWM sources,[7, 10, 11] the integration of energy tunable quantum light sources on silicon is arguably one of the most promising choices for miniaturization and on-chip quantum information applications. Despite these pressing needs, the experimental demonstration is still missing.

Here we demonstrate, for the first time, energy tunable entangled photon sources based on III-V QDs integrated on a silicon chip. It has been predicted that the FSS of QDs can be effectively eliminated by uniaxial stresses when the strain axis is closely aligned along the [110] or [1-10] direction.[28, 29] This finding has been experimentally verified by our group in a recent work,[22] and a high percentage of QDs can be tuned into entangled photon sources with this method. Although it removes the tedious search for the "hero" QDs, the energy of entangled photons is not yet deterministic and also only *1 QD per chip* can be tuned. These are, unfortunately, the common disadvantages associated with all "macroscopic" FSS tuning technologies to date. To this end, we design and fabricate a device consisting of QD-embedded nanomembranes suspended on a four-legged thin-film PMN-PT ($[Pb(Mg_{1/3}Nb_{2/3})O_3]_{0.72}[PbTiO_3]_{0.28}$) actuator integrated on a silicon substrate. With the combined uniaxial stresses along two orthogonal directions, we are able to keep the FSS *strictly* below 1 μeV while shifting the exciton energy by more than 3000 times of the QD radiative linewidth. High fidelity entangled photon emission is demonstrated when the FSS is tuned to below 1 μeV. Therefore energy tunable entangled photons are generated on chip with a single device footprint of a few hundred microns.

For the device fabrication we use the industrial transfer printing and die bonding techniques to realize the integration of III-V, PMN-PT and Si. Unlike the thick PMN-PT substrate used in all previous works,[21, 22, 28, 30, 31, 32] a 15 μm PMN-PT thin film bonded on a silicon substrate is employed here. This PMNPT-on-Si technology has been studied extensively in recent years, and we use it to realize the scalable integration of MEMS with sophisticated functionalities on chip (Fig. 1, see also *Methods*). Arrays of QD-containing GaAs nanomembranes, each 80×80 μm$^2$ in size, were then transferred onto the PMN-PT MEMS with four actuation legs (Fig. 1). The crystal axes [1-10] and [110] of the GaAs nanomembrane were carefully aligned along the designed stress axes of the actuators. When applying negative (positive) voltages to the electric contacts, the PMN-PT legs expand (contract) in-plane and therefore exert quasi-uniaxial compressive (tensile) stresses to the bonded QD nanomembranes.

This new device concept has several advantages. First, a controllable anisotropic strain is achieved by the four legged configuration which is not possible with the bulk PMN-PT. Second, the use of piezoelectric film alleviates us from high voltages (typically, up to thousands volts) requried for the bulk PMN-PT, which is certainly important for on-chip integration. And third, theoretically proposed scalable sources of *strain controlled* quantum light sources[32, 33] can be realized on chip with such devices.

By sweeping the voltage on *only* one pair of the legs (*e.g.* $V_{BD}$ on legs B&D), the exciton emission energy is shifted over a large range (up to 12 meV in this work, see Fig. 1d) with a rate of 50 to 100 µeV/V for different devices. For suitable QDs the exciton emission can be even tuned across the Cesium D1 line at 1.3859 eV, see Fig. 1d, which is required for realizing a hybrid quantum memory.[34] During the wavelength tuning, the FSS changes monotonously (not shown here) which reflects the quasi-uniaxial nature of the applied stress. With realistic parameters for PMN-PT, GaAs and Si, we have performed finite element time domain (FDTD) simulations of our device and two principle stresses (with a magnitude of about 2 GPa at voltages of 50 V) along orthogonal directions can be identified (not shown here). Similar to previous findings,[28, 29, 31] the uniaxial strain tuning of FSS is determined by the QD principal axis[14] with respect to the stress direction. For a QD whose principal axis is closely aligned with the stress direction, the FSS can be effectively eliminated and the polarization direction $\theta$ (or, say, phase) of the high-energy component of the exciton undergoes a sharp phase change of 90 degrees. For a QD whose principal axis is misaligned with the stress direction, the tuning of FSS has a lower bound, and the polarization direction gradually rotates with a phase change of less than 90 degrees.[29]

How does a QD behave when both voltages $V_{AC}$ and $V_{BD}$ are turned on? This has never been studied, due the lack of realistic experimental techniques. In Fig. 2 we discuss the behavior of a QD under the application of a pair of orthogonal uniaxial stresses. The directions of the principal axes of the epitaxial semiconductor QDs experience a Gaussian-like distribution, and for a significant amount of QDs the exciton polarizations are closely aligned with the preferred crystallographic directions. Our recent statistic study show that up to 30% of the QDs can be tuned to emit polarization entangled photons, by uniaxial stress along either [110] or [1-10] directions.[22] In our case this can be done by sweeping the voltage, *e.g.* $V_{AC}$, on one pair of the actuation legs, while fixing the voltage $V_{BD}$ on the other pair of legs. In Fig. 2a and b we fixed $V_{BD}$ at 0 and -25 V, respectively. We observe that the polarization of this QD, at the initial voltage combination ($V_{AC}$, $V_{BD}$) of (0, 0) V, is accurately aligned with respect to the [1-10] direction, or, equivalently, parallel to the legs A&C (see Fig. 2a). Here the phase $\theta$ indicates the angle between the exciton polarization and the [1-10] crystallographic direction of GaAs, see inset. Both Fig. 2a and b reproduce fully the theoretical predictions,[28, 29] *i.e.* a minimum FSS at around 0 μeV and an abrupt change in $\theta$ by *exactly* 90 degrees. The only difference between the two situations ($V_{BD}$ = 0 and -25 V) is the voltage $V_{AC}$ at which the FSS reaches the minimum.

Two dimensional scanning on the two pairs of legs, by sweeping both $V_{BD}$ and $V_{AC}$ is then performed. In Fig. 2c we show the results in a three dimensional plot. The astonishing result is that, with this four-legged device providing orthogonal uniaxial stresses, multiple zero FSS points with different exciton energy $E_X$ can be achieved. At different $V_{BD}$, the electronic symmetry of quantum dot can be always recovered by sweeping $V_{AC}$ and the FSS is erased, see also Fig. 4a. The dashed line on the bottom plane of the plot indicates the combinations of ($V_{AC}$, $V_{BD}$) at which the FSS reaches its minimum. A linear relationship is found and the ratio of voltage changes $\Delta V_{AC}/\Delta V_{BD}$ is about 0.58. In terms of the applied stresses (*X, Y*), indeed, an effective two-level model for the FSS of QDs with exciton polarization closely aligned to principal stress axes would predict a zero FSS with a linear relationship $\Delta X/\Delta Y$ of 1, see Fig. 4b.

Therefore the ratio of 0.58 for the voltage changes is attributed to the in-plane anisotropic strain properties of the PMN-PT actuation legs. To the best of our knowledge, this high degree of control on FSS has never been realized with strain,[32] and will be extremely difficult to implement by other technique developed prior to this work.

We also investigate the changes in exciton energies and results are given in Fig. 2d. With negative voltages applied, the legs exert uniaxial compressive stresses to the QD which causes a blue-shift in exciton energy. By sweeping $V_{AC}$ at a fixed $V_{BD}$, we observe clear monotonic changes in the exciton energy of a few meV. The effect should be the same when sweeping $V_{BD}$ at a fixed $V_{AC}$. The effect of tensile stresses in one direction is almost identical to that of compressive stresses in the orthogonal direction.[33] In Fig. 2d the data on the left vertical plane are taken at $V_{AC}$ = 50 V and the dashed line indicates a clear blue-shift when sweeping $V_{BD}$ from 25 to -100 V, i.e. from the tensile region to the compressive region. Therefore we also have a high degree of control on the exciton energies by using two actuation legs.

Since the independent tunability of exciton energy and FSS is a main concern in this work, we plot in Fig. 3a the FSS *versus* exciton energies for different $V_{BD}$. For clarity, we show only the FSS range from 0 to 5 µeV. It shows exactly a linear behavior (represented by the solid line fits) in agreement with a **k.p** analysis, see a later discussion. For self-assembled InAs/GaAs QDs the relationship between the FSS and the attainable entanglement fidelity $f^+$ has been well documented and it is commonly accepted that the entanglement persists for a FSS even up to 3-4 µeV.[21, 22, 35] As shown in these reports, tuning the FSS to less than 1µeV yields a highly fidelity $f^+$ of larger than 0.7. With our new device, it is very clear that the FSS can be "locked" *strictly* below 1 µeV, *i.e.* high fidelity entangled photons can be generated from the QDs, for a large range of exciton energies. Considering the typical lifetime (500 *ps* to 1 *ns*) in our QDs and therefore a radiative limited linewidth of about 1 µeV, the tuning range of 3.7 meV shown in Fig. 3a corresponds to more than 3000 times of the radiative linewidth. This tunability is more than one order of magnitude larger than what has been demonstrated for SPDC sources.[6]

We have performed the polarization cross-correlation spectroscopy[17, 20, 21, 23] on a brighter QD embedded inside another device *on the same chip*. The FSS is tuned to around zero (0.21 ± 0.20 µeV) to demonstrate the polarization entanglement, and the data are presented in Fig. 3b. A key criterion for entanglement is the presence of a correlation independent of the chosen polarization basis, *i.e.* $|\Psi^+> = 1/\sqrt{2}\,(|H_{XX}H_X> + |V_{XX}V_X>) = 1/\sqrt{2}\,(|D_{XX}D_X> + |A_{XX}A_X>) = 1/\sqrt{2}\,(|R_{XX}L_X> + |L_{XX}R_X>)$, with D, A, R, L denoting the diagonal, anti-diagonal, right-hand circular and left-hand circular polarizations. Clear photon bunching, with a normalized second order correlation function $g^2(\tau) > 3$, can be observed for the co-polarized HH and DD photons, whereas in the circular basis the bunching occurs for the cross-polarized RL photons. The entanglement fidelity $f^+$ to the maximally entangled Bell state can be determined from the measurements in Fig. 3b, see *Methods*. The peak near the zero time delay yields a fidelity $f^+$ of 0.733 ± 0.075 without any background subtraction, which exceeds the threshold of 0.5 for a classically correlated state by more than 3 standard deviations. The above results are in line with previous experimental and theoretical works, and verify that highly entangled photons can be generated with our device with large energy tunabilities.

An intuitive understanding of the tuning behavior can be obtained immediately from a matrix of exciton polarization plots at different voltage combinations, see Fig. 4a. The "circularity" of the polarization pedals indicates the relative magnitude of FSS. By sweeping $V_{AC}$ from 0 to 70 V at $V_{BD}$ of 0V, an increasing tensile stress is applied along legs A&C and the symmetry is gradually recovered. We observe the "opening" of the polarization pedal without any appreciable rotation,[19, 20, 31] which is in perfect agreement with theoretical predictions.[29] At the voltage combination ($V_{AC}$, $V_{BD}$) = (70, 0) V, the exciton emissions become circular polarized with a near-zero FSS and polarization entangled photons are generated. At $V_{BD}$ of -25 V and -50 V, the symmetry is already partially recovered due to the increased compressive stress along the legs B&D, and therefore the symmetry can be fully recovered with less tensile stress, *i.e.* a smaller $V_{AC}$, along the legs A&C. Due to the symmetry of orthogonal uniaxial stresses, we can observe the same effect when sweeping $V_{AC}$ at fixed voltages of $V_{BD}$.

Our experimental findings can also be qualitatively understood using the recent theoretical observation[31] according to which the application of two orthogonal stresses of magnitude $X$ and $Y$ can be used as effective knobs to erase the FSS and tune the exciton emission once the stress axis are aligned with the natural polarization direction of the excitons in the absence of strain effects. In this case indeed, the FSS scales linearly with the stress anisotropy $\Delta = X - Y$ and eventually vanishes at a critical stress anisotropy $\Delta = \Delta c$. This behavior is demonstrated in Fig. 4b where we show a density plot of the FSS as a function of the two stress magnitudes $X, Y$ as obtained with the two-level model Hamiltonian[29, 31]. Furthermore, the remaining independent hydrostatic part $\propto X + Y$ can be used to change at will the exciton energy $E_X$. This, in turn, implies a linear interdependence between the exciton emission and the FSS for an aligned QD, which is in perfect agreement with the data presented in Fig. 2.

We emphasize that, for a number of reasons, the device presented here holds strong promise for the realization of scalable quantum information processing. *First*, the energy tunability allows the Bell state measurement between two such sources and therefore the swapping of entanglement based on a solid-state platform. Also, with suitable QDs (see for example Fig. 1d) and improved photon collection efficiencies, it is straightforward to couple the source with atomic vapors. *Second*, the integrated MEMS on silicon have small footprints and low operation voltages, which solves two of the most challenging problems of the strain engineering technique.[21,30] We can foresee the integration of this device with other photonic structures and, most intriguingly, with advanced silicon quantum photonic circuits. The envisioned hybrid can exploit the maturity of CMOS technology, and as well as the waveguiding, processing and detection capabilities associated with silicon photonics. Third, a significant amount of QDs (up to 30% of studied QDs) can be found as ideal candidates for an energy tunable entangled photon source, which removes the tedious search for "hero" QDs. As proposed in a recent theoretical work,[32] a six-legged device based on a *bulk* PMN-PT substrate can be used *in principle* to tune all available QDs, and the implementation of such a design is possible with our devices.

In summary, we have experimentally realized energy tunable entangled photon sources on a III-V/Si chip. This hybrid photonic device holds strong promise for the realization of scalable deterministic quantum light sources based on III-V QDs. The MEMS based device features the advantages of easy operation and on-chip integration. We envision that it will inspire many other photonic devices, such as photonic crystal circuits, where the on-chip engineering of quantum light sources is essential. With further improvements on photon collection efficiency, entanglement fidelity and photon coherences,[25,27] this device will play an important role not only in building a solid-state quantum network based on entanglement swapping and quantum memories, but also in building advanced quantum photonic circuits for on-chip information processing.

# Method

## Sample growth

The studied sample was grown on a (001) GaAs substrate by solid-source molecular beam epitaxy (MBE). Several samples were used in this work, but their general structures are the same. After the deoxidization and GaAs buffer growth, a thin layer of $Al_{0.75}Ga_{0.25}As$ was grown as the sacrificial layer for the nanomembrane release. The InAs QDs were grown by partial capping and annealing, and embedded in the middle of a GaAs layer with a thickness of a few hundred nanometers. The emission range of the QDs is between 880 nm and 900 nm and we can easily identify low density regions with less than 1 QD per $\mu m^2$.

## Device fabrication

As for the device processing, we start with the QD-embedded nanomembranes. Standard UV photolithography and wet chemical etching were used to fabricate mesa structures with a size of $80\times 80$ $\mu m^2$. The edge of the nanomembranes was processed along [110] or [1-10] crystal axis of GaAs. The PMN-PT film is around 15 micron thick. The backside is coated with gold as the bottom/ground contact. And the PMN-PT film is bonded to a silicon substrate via glue (Ibule photonics). We use focused ion beam to define the trenches in the film, and the depth of the trenches is deep enough to penetrate into the silicon substrate. And then the top contact is deposited by E-beam sputtering. After this, the device is undercut by wetting chemical etching to form the suspended PMN-PT thin film legs.

Then we bond the QD-embedded nanomembranes to the processed PMN-PT/Si substrate using a flip-chip bonder. The edge of nanomembrane is carefully aligned along the *strain* axis of the PMN-PT actuation legs. In order to constrain the displacements of the membrane, the same voltage is applied to opposite legs.

**Optical measurement**

For optical measurements, the device is loaded into a cryostat chamber which is cooled to around 5K. The PL emitted from the QDs is collected by a 100× microscope objective with numerical aperture of 0.52. The signal is then dispersed by a spectrometer with 1800 grating and detected by a liquid nitrogen cooled CCD. By inserting a half-wave plate and a linear polarizer directly after the collection lens, polarization-resolved measurements were performed to estimate the FSS. The exciton polarization is determined by aligning the fast optical axis of the polarizer along [1-10] direction of the nanomembrane. With the experimental procedure used in previous works,[19,31] we can determine the FSS with an accuracy of sub-µeV. As for the polarization correlation spectroscopy, a non-polarizing 50:50 beam splitter is placed directly after the objective in order to divide the optical paths between two spectrometers, which are then used to detect X and XX separately. After each spectrometer, a Hanbury-Brown Twiss setup, consisting of a polarizing beam splitter and two high efficiency single-photon avalanche detectors, is placed. Half- and quarter-wave were used to select the proper polarization basis. The temporal resolution of the system is about 450 ps. The entanglement can be quantified by measuring degree of correlation C, which is defined by

$$C_{basis} = (g^{(2)}_{XX,X}(\tau) - g^{(2)}_{XX,\bar{X}}(\tau))/(g^{(2)}_{XX,X}(\tau) + g^{(2)}_{XX,\bar{X}}(\tau)),$$

Where $g^{(2)}_{XX,X}(\tau)$ and $g^{(2)}_{XX,\bar{X}}(\tau)$ are normalized second-order time correlations for co-polarized and cross-polarized XX and X photons, respectively. The fidelity $f^+$ is calculated by using the formula: $f^+ = (1 + C_{HV} + C_{DA} - C_{RL})/4$, in which $C_{HV}, C_{DA}$, and $C_{RL}$ are degree of correlations in HV, DA and RL bases.

## ACKNOWLEDGEMENT

The work was financially supported by the BMBF QuaHL-Rep (Contract No. 01BQ1032 and 01BQ1034), Q.Com-H (16KIS0106) and the European Union Seventh Framework Programme 209 (FP7/2007-2013) under Grant Agreement No. 601126 210 (HANAS). The authors acknowledge Carmine Ortix for theoretical supports. Part of the data analysis was done by the XRSP3 software developed by Armando Rastelli. We thank Yongheng Huo for providing one of the test QD sample materials. Some of the samples were grown by R. Keil for this work. We also thank B. Höfer, C. Jiang, B. Eichler, R. Engelhard, M. Bauer and S. Harazim for discussions and technical supports. F. D. acknowledges Armando Rastelli for the continuous support and his early contributions to this project.

## AUTHOR CONTRIBUTIONS

F. D. conceived the experiment and supervised the project together with O. G. S who directed the research. The devices were designed by Y. C. and F. D., and fabricated by Y. C. with the help from K. Jung and Y. Z. Optical experiments were performed by Y. C. and J. Z. The data were analyzed by Y. C., F. D. and J. Z. FDTD simulations were done by M. Z and Y. Z. The manuscript was written by Y. C. and F. D., with the inputs from all the authors.

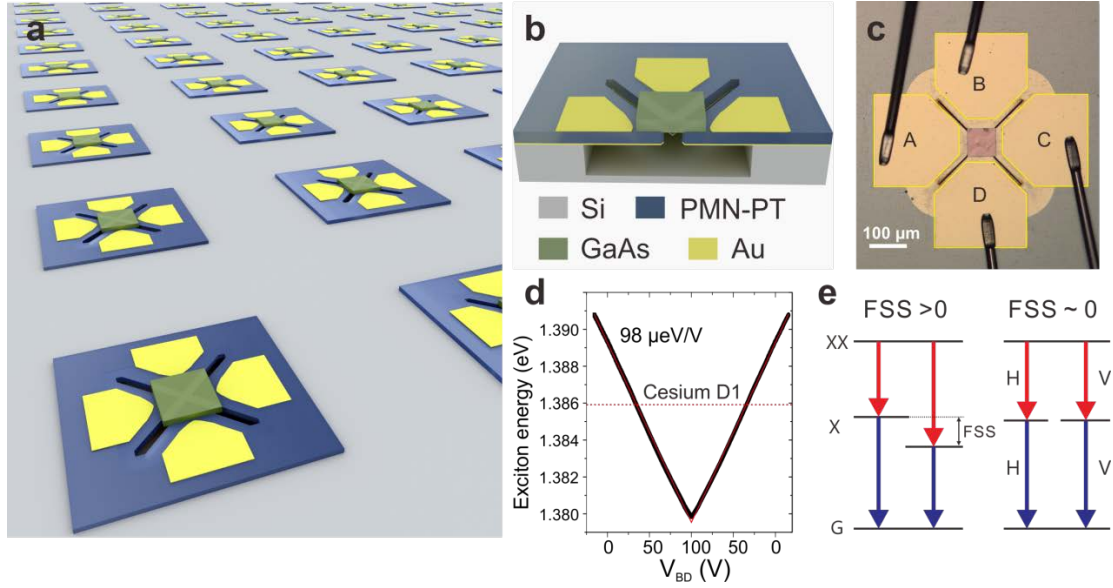

**Figure 1. Energy-tunable polarization entangled photon sources integrated on silicon.** (a) MEMS devices for anisotropic strain engineering of III-V QD based quantum light sources. Due to its small footprint and the compatibility with mature semiconductor technologies, large scale on-chip integration is feasible. (b) Schematic of the cross section of a single device. FIB cut is used to define trenches on the PMN-PT thin film, and then wet chemical undercut is used to remove the underlying Si and to form four actuation legs. A thin GaAs nanomembrane containing InAs QDs is transferred onto the suspended region between the four legs. (c) Micrograph showing the zoom-in of a completed device. Electrical contacts are made on the four legs A-D. The center region is a bonded QD-containing nanomembrane. (d) Performance of a typical device. The exciton energy of a single QD is recorded when the voltage on legs B&D is scanned. The actuation legs contract under positive voltages, leading to the tensile stresses on the nanomembrane and to the red-shift of the QD emission. The red solid lines show the linear fit. The dashed line indicates that the energy of this QD can be scanned across the cesium D1 absorption line. (e) Illustration of the fine structure splitting (FSS) in a QD. Polarization entangled photons are emitted from the XX cascade only when the FSS is tuned to near zero.

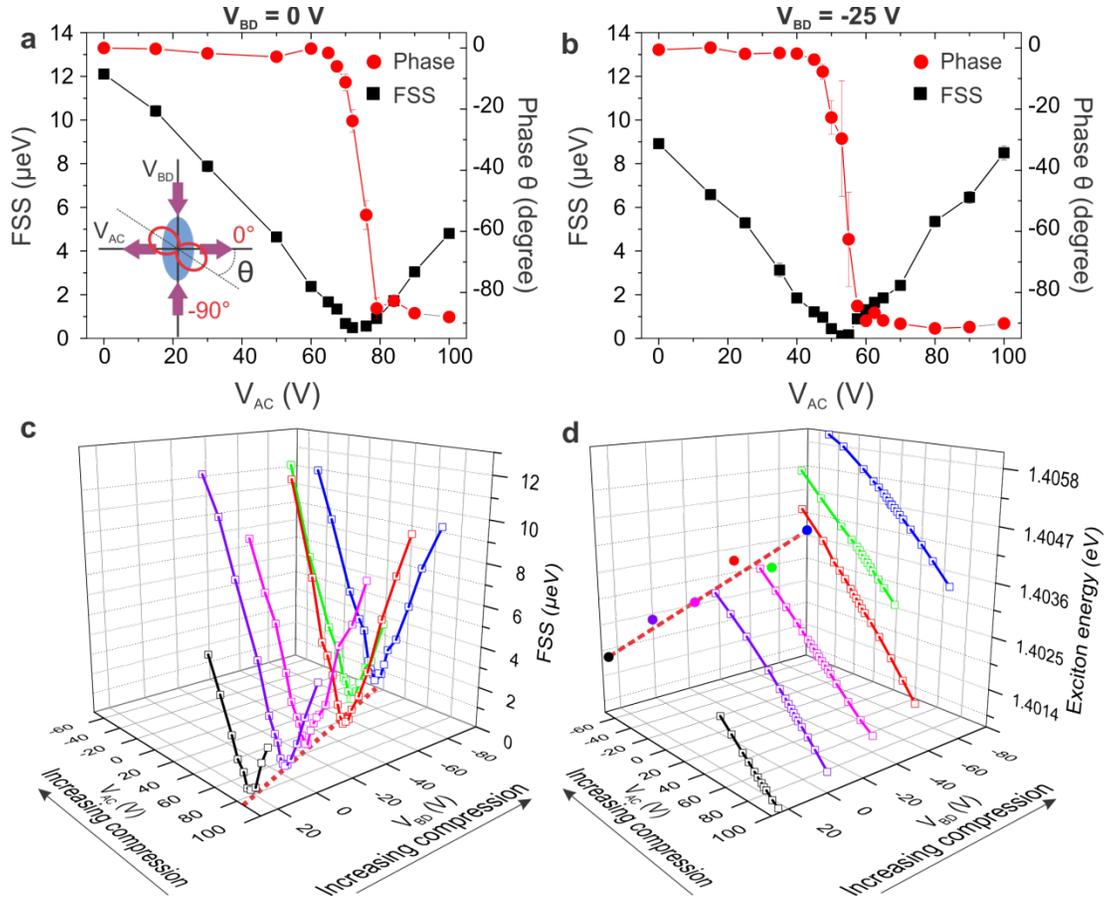

**Figure 2. Anisotropic strain engineering of a QD under orthogonal uniaxial stresses.** (a), (b) FSS and phase θ are plotted as a function of the voltage $V_{AC}$ at a fixed voltage $V_{BD}$ of 0 V and -25 V, respectively. With the increasing voltages on legs A&C, FSS decreases monotonically to around zero and then increases again. And the phase shows an abrupt change from 0 to -90 degree, when FSS reaches the minimum value. The inset in (a) gives the definition of θ. The ellipse indicates an elongated QD with its major axis aligned along a crystallographic direction. The red solid line indicates the exciton polarization. (c) The changes in FSS when both $V_{BD}$ and $V_{AC}$ are scanned. The dashed line on the bottom plane indicates a linear shift of the voltage combination ($V_{AC}$, $V_{AC}$) at which FSS reach the minimum values. (d) During the scanning of $V_{BD}$ and $V_{AC}$, the exciton energies are recorded. The data points on the left vertical plane are taken at $V_{AC}$ of 50 V, and the dashed line is a linear fit.

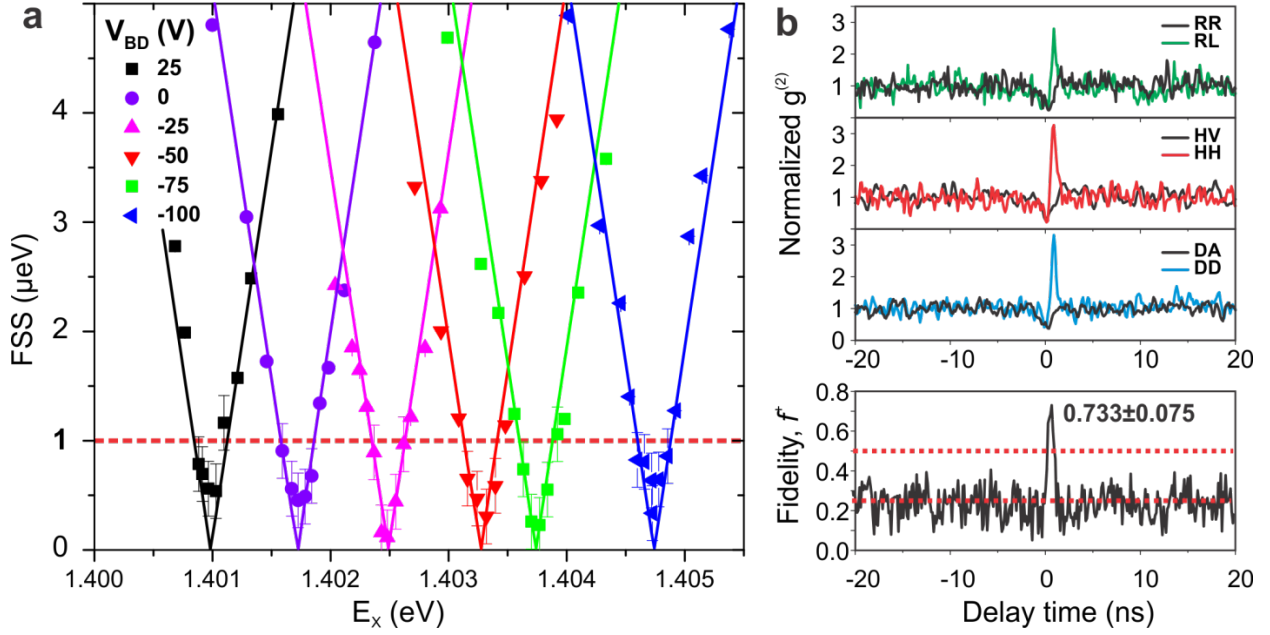

**Figure 3. Independent tunability of exciton energies and FSS.** (a) FSS is plotted as a function of the exciton energy $E_X$, at different values of $V_{BD}$. The solid lines are theoretical fits. In the **k·p** theory we consider the effect of a pair of orthogonal uniaxial stresses applied to a QD, when its initial polarization direction is aligned with one of the crystallographic directions. Exciton energy at which FSS ~ 0 is tuned by 3.7 meV. The dashed line is a threshold of 1μeV for the entangled photons generations. (b) Polarization correlation spectroscopy, see *Methods*, are performed on the biexciton and exciton photons, when the QD FSS is tuned to a minimum value of 0.21 ± 0.20 μeV. The normalized coincident counts are given for both co-polarized and cross-polarized photons. We have measured a fidelity $f^+$ of 0.733 ± 0.075 without any background subtraction. The two dashed lines indicate the threshold of 0.5 for the classically correlated light, and the threshold of 0.25 for the uncorrelated light.

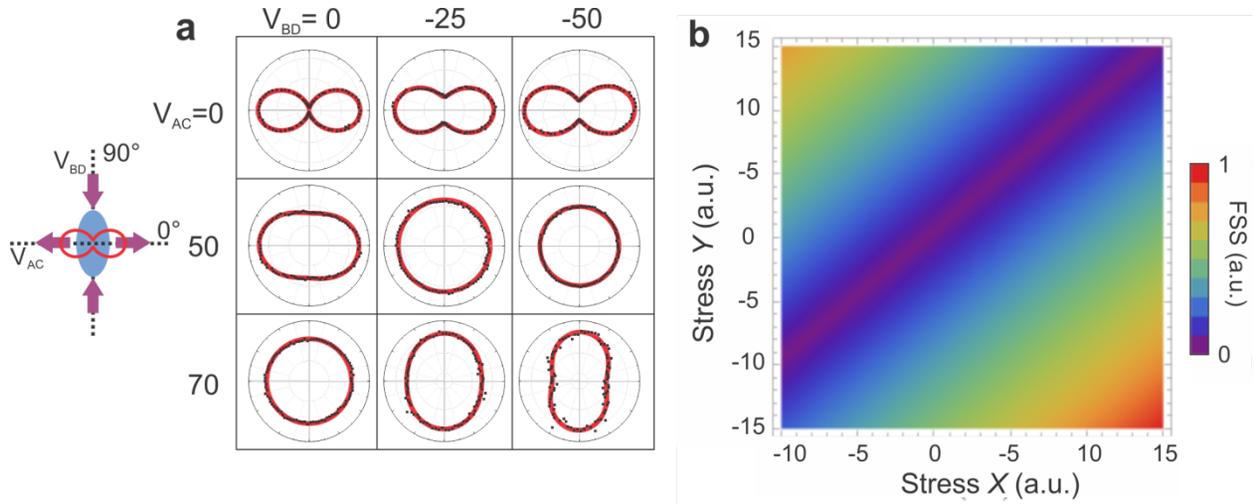

**Figure 4.** (a) Matrix of polarization plots at different voltage combinations. Here the exciton energy is plotted as a function of $\theta$, with the same definition being used in the inset of Fig. 2a. All polar plots have an axis scale of 15 µeV, and the solid red lines represent a fit to the data with a sinusoidal function. The "circularity" and the direction of the pedals indicate the relative amplitude of FSS and $\theta$, respectively. Circular polarized exciton emission with near zero FSS can be observed at three different voltage combinations, *i.e.* at different exciton emission energies. No appreciable polarization rotation can be observed at the applied voltages. The schematic shows the stress condition for this type of QDs. (b) A density plot of the FSS as a function of the two stress magnitudes $X, Y$ as obtained with the two-level model Hamiltonian.